\documentclass[conference,a4paper]{IEEEtran}
\usepackage{xcolor}
\usepackage{balance}
\usepackage{placeins}

\usepackage{amsmath}
\usepackage{stix}

\ifCLASSINFOpdf
  \usepackage[pdftex]{graphicx}
\else
  \usepackage[dvips]{graphicx}
\fi
\ifCLASSOPTIONcompsoc
 \usepackage[caption=false,font=normalsize,labelfont=sf,textfont=sf]{subfig}
\else
 \usepackage[caption=false,font=footnotesize]{subfig}
\fi
\begin{document}
\twocolumn[
\begin{@twocolumnfalse}

\textbf{Author's pre-print} \\

\copyright 2021 IEEE. Personal use of this material is permitted. Permission from IEEE must be obtained for all other users, including reprinting/ republishing this material for advertising or promotional purposes, creating new collective works for resale or redistribution to servers or lists, or reuse of any copyrighted components of this work in other works.\\


\end{@twocolumnfalse}
]

\newpage
%
\title{Antenna De-Embedding in FDTD Using Spherical Wave Functions by Exploiting Orthogonality}

\author{\IEEEauthorblockN{
Leonardo M\"orlein\IEEEauthorrefmark{1},   
Lukas Berkelmann\IEEEauthorrefmark{1},   
Dirk Manteuffel\IEEEauthorrefmark{1}   
}                                     
\IEEEauthorblockA{\IEEEauthorrefmark{1}
Institute of Microwave and Wireless Systems, Leibniz University Hannover, \\
Hannover, Germany, \{moerlein,manteuffel\}@imw.uni-hannover.de
}
}



\maketitle

\begin{abstract}
De-embedding antennas from the channel using Spherical Wave Functions (SWF) is a useful method to reduce the numerical effort in the simulation of wearable antennas. In this paper an analytical solution to the De-embedding problem is presented in form of surface integrals. This new integral solution is helpful on a theoretical level to derive insights and is also well suited for implementation in Finite Difference Time Domain (FDTD) numerical software. The spherical wave function coefficients are calculated directly from near-field values. Furthermore, the presence of a near-field scatterer in the de-embedding problem is discussed on a theoretical level based on the Huygens Equivalence Theorem. This makes it possible to exploit the degrees of freedom in such a way that it is sufficient to only use out-going spherical wave functions and still obtain correct results.
\end{abstract}

\vskip0.5\baselineskip

%

\section{Introduction}

While separating the antenna from the channel is straightforward in free-space scenarios, antenna de-embedding in wireless body area networks is complex \cite{BodyCentricWireless}. To reduce the simulation domain during antenna design, antenna de-embedding using SWF has been proven to be a helpful tool \cite{Naganawa2015}, \cite{Naganawa2017}. Instead of simulating the antenna in the full propagation scenario, the channel is pre-simulated and a linear mapping in form of the channel matrix $\mathbf{M}$
\begin{equation}
    \mathbf{a}_\mathrm{R} = \mathbf{M} \, \mathbf{b}_\mathrm{T},
\end{equation}
is obtained, which relates out-going SWF coefficient vectors at the transmitting antenna $\mathbf{b}_\mathrm{T}$ to in-coming SWF coefficient vectors at the receiving antenna $\mathbf{a}_\mathrm{R}$. The channel is therefore excited up to a certain order by single mode spherical wave sources one after the other.
The antenna itself is then simulated in a smaller simulation domain only containing the tissue, which directly affects the current distribution on the antenna itself. The near-field results from this simulation are then decomposed into spherical waves to obtain the specific $\mathbf{b}_\mathrm{T}$ for the actual antenna. The channel matrix $\mathbf{M}$ is then used to calculate the SWF coefficients $\mathbf{a}_\mathrm{R}$ at the receiving antenna, which is then translated into a link-budget for a specific receiving antenna. Iteration during the antenna design procedure is therefore possible without resimulating the entire link scenario.


While the channel modeling approach to obtain $\mathbf{M}$ is well described in the work of Naganawa et al. \cite{Naganawa2017}, an alternative approach to obtain the $\mathbf{b}_\mathrm{T}$ that offers some advantages is proposed in this work. To avoid in-coming spherical waves, Naganawa et al. suggested a complex procedure to decompose the near-fields into spherical wave coefficients. Their approach involves an additional step to calculate the electromagnetic field caused by the on-body antenna current distribution displaced into free-space. Understanding and simplifying this method in terms of complexity is the main motivation for this paper. The following sections show that the decomposition is possible directly from the recorded on-body near-field values.

In this paper, the following questions will be examined therefore:
\begin{enumerate}
    \item Is there a direct method to solve the SWF decomposition problem in favor of the indirect least mean squares / pseudo-inverse approach they presented?
    \item Is there a more simple decomposition approach (avoiding the current displacement) that is still suitable for this kind of application?
\end{enumerate}

\section{Mathematical Decomposition of the Fields}

\begin{figure}
    \centering
    \includegraphics{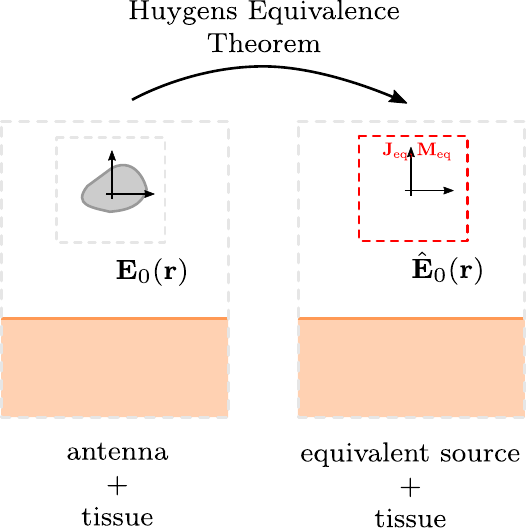}
    \caption{Sketch illustrating the Huygens Equivalence Theorem}
    \label{fig:equivalence}
\end{figure}

While Naganawa et al. only use a pseudo-inverse for their decomposition \cite{Naganawa2015}, \cite{Naganawa2017}, this paper proposes an orthogonality based decomposition technique, that is analytically explained in this section.

According to Hansen \cite{Hansen1988}, an electrical field in free-space can be seen as a linear combination of spherical wave functions at an arbitrary origin point:
\begin{equation} \label{eq_e_total_sum}
    \mathbf{E} = k \sqrt{\eta} \sum_j b_j \mathbf{F}_j^{(4)} + a_j \mathbf{F}_j^{(3)},
\end{equation}
whereby $k$ is the wave number, and $\eta$ is the wave impedance in free-space. In contrast to Hansen, we assume the time-dependence $e^{j\omega t}$, so $\mathbf{F}_j^{(4)}$ and $b_j$ represent out-going SWF and $\mathbf{F}_j^{(3)}$ and $a_j$ represent inwards travelling SWF, while the mathematical formulas for the SWF  $\mathbf{F}_j^{(c)}$ with the mode type index $c \in \{1,3,4\}$ remain as defined by Hansen.

The total electrical field $\mathbf{E}$ can be decomposed into the components using the mathematical orthogonality relations of spherical wave functions. Using an arbitrarily shaped, closed surface $S$ around the origin point of the SWF, the notation
\begin{equation}
    \label{eq_integral_definition}
    \left<\, \mathbf{u},\; \mathbf{v}\, \right> =
    \oiint_S \; \left\{
          \mathbf{u} \times \left( \nabla \times \mathbf{v} \right)
        - \mathbf{v} \times \left( \nabla \times \mathbf{u} \right)
    \right\} \mathbf{\hat{n}} \; \mathrm{d}S
\end{equation}
and by combining the derivations of Hansen \cite[pp.330]{Hansen1988} and Kristensson \cite[pp.40]{Kristensson}, the orthogonality relation
\begin{equation}
\label{eq_swf_orthogonality_end}
    \left<\, \mathbf{F}_j^{(c)},\; \mathbf{F}_{j'}^{(\xi')*} \right> = \frac{1}{2\,k^2} \delta_{jj'}\, B^{(c,\xi')},
\end{equation}
with $\delta_{jj'}$ as Kronecker delta and $B^{(c,\xi')}$ from Table~\ref{tab:a_cc} can be derived. This relation can be used to perform the decomposition of the total field $\mathbf{E}$ from \eqref{eq_e_total_sum} into the spherical wave function coefficients:
\begin{equation}
    \label{eq_d4_operator}
    b_j = \frac{k}{j\sqrt{\eta}} \left<\mathbf{E}, \mathbf{F}_j^{(4)*}\right>,
\end{equation}
\begin{equation}
    \label{eq_d3_operator}
    a_j = -\frac{k}{j\sqrt{\eta}} \left<\mathbf{E}, \mathbf{F}_j^{(3)*}\right>.
\end{equation}
While this condensed mathematical formulation is useful for analytical proofs, the term $\nabla \times \mathbf{E}$ in the inner product \eqref{eq_integral_definition} can be expressed in terms of $\mathbf{H}$, which leads to variants of \eqref{eq_d4_operator} and \eqref{eq_d3_operator} more suitable to calculate coefficients directly from FDTD simulation results. One of the major benefits of \eqref{eq_d4_operator} and \eqref{eq_d3_operator} is, that the surface $S$ can be arbitrarily shaped (around the reference origin). This way, the integral can be implemented by a summation of the electromagnetic field values on the surface of a box aligned with the grid directions in rectangular FDTD schemes.

For convenience, the coefficients $b_j$ and $a_j$ are written as coefficient vectors $\mathbf{b}$ and $\mathbf{a}$ throughout the paper, whenever suitable.

\section{Antenna De-embedding in Presence of a Backscatterer}

The antenna dembedding scheme presented in this paper is derived following the Huygens Equivalence Theorem \cite[pp.328]{Balanis2012} as depicted in Fig.~\ref{fig:equivalence}. According to the theorem the radiating antenna is removed from the scenario and replaced by equivalent currents, e.g. $\mathbf{J}_{\mathrm{eq}} = \hat{\mathbf{n}} \times \mathbf{H}_0 $ and $\mathbf{M}_{\mathrm{eq}} = -\hat{\mathbf{n}} \times \mathbf{E}_0 $, on an enclosing surface. In the case of rectangular FDTD schemes, a box is the most suitable circumscribing surface so the cartesian coordinates are aligned with the surfaces.

The Huygens Equivalence Theorem relies on the idea, that the electromagnetic field outside of a certain volume $V$ should remain constant. When the desired equivalence is fulfilled, then 
\begin{equation}\label{eq_equivalence}
    \mathbf{\hat{E}_0}(\mathbf{r}) = \mathbf{E_0} (\mathbf{r})  \hspace{0.5cm}\mathrm{for}\hspace{0.5cm} \mathbf{r}\hspace{0.1cm}\text{outside of}\hspace{0.1cm}V
\end{equation}
is valid in Fig.~\ref{fig:equivalence}. The solution $\mathbf{J}_{\mathrm{eq}} = \hat{\mathbf{n}} \times \mathbf{H}_0 $ and $\mathbf{M}_{\mathrm{eq}} = -\hat{\mathbf{n}} \times \mathbf{E}_0 $ is only one solution to the problem, leading to
\begin{equation}
    \hspace{0.4cm}\mathbf{\hat{E}_0}(\mathbf{r}) = \mathbf{0}  \hspace{0.7cm}\mathrm{for}\hspace{0.5cm} \mathbf{r}\hspace{0.1cm}\text{inside of}\hspace{0.1cm}V.
\end{equation}
This special solution is called Love's Equivalence Theorem. However, the Huygens Equivalence Theorem includes a lot more solutions with different inner electromagnetic fields. As the de-embedding problem has no constraint on the inner electromagnetic fields, all solutions are of interest here.  




\begin{table}[]
    \renewcommand{\arraystretch}{1.5} 
    \centering
    \[
    \begin{array}{ccccc}
    \hline B^{(c, \xi')} & {\xi'=1} & {\xi'=4} & {\xi'=3} \\ \hline
    c=1 & {0} & {j} & {-j} \\
    {c=3} & {-j} & {0} & {-2 j} \\
    {c=4} & {j} & {2 j} & {0} \\ \hline
    \end{array}
    \]
    \caption{Orthogonality coefficient $B^{(c, \xi')}$ between different radial SWF dependencies similar to $A^{(c, c')}$ from \cite[p.315]{Hansen1988}.}
    \label{tab:a_cc}
\end{table}

\begin{figure}
    \centering
    \includegraphics{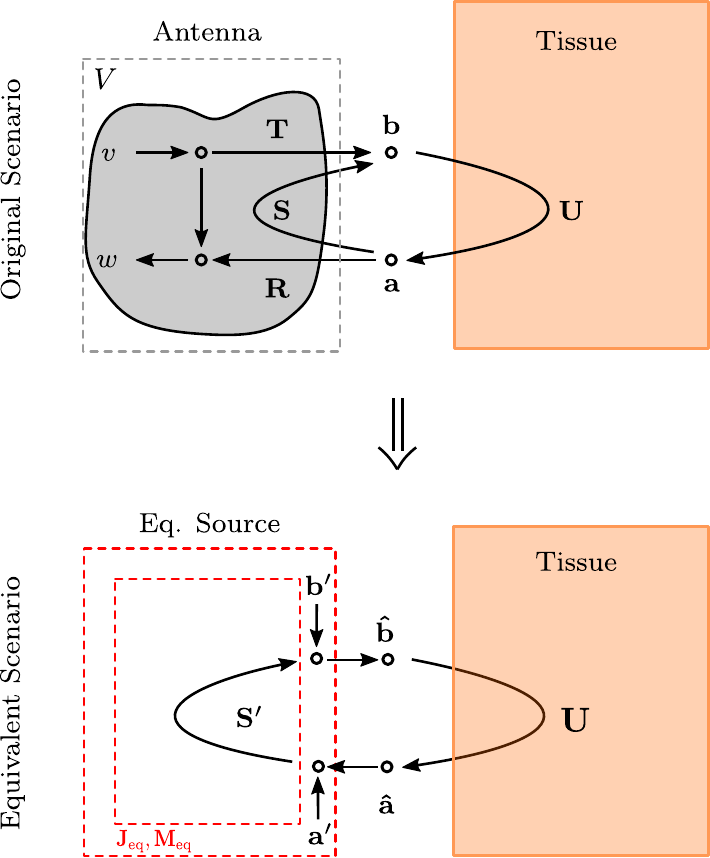}
    \caption{Signal flow graph representation of the original problem (upper part) and signal flow graph representation with the equivalent source (lower lart).}
    \label{fig:antenna_refl_screnarios}
\end{figure}

The original field $\mathbf{E}_0$ around the antenna consists of in-coming and out-going waves:
\begin{equation} \label{eq:e0_original}
    \mathbf{E}_0 = k \sqrt{\eta} \sum_j b_j \mathbf{F}_j^{(4)} + a_j \mathbf{F}_j^{(3)},
\end{equation}
weighted by the coefficients $b_j$ and $a_j$. The upper part of Fig.~\ref{fig:antenna_refl_screnarios} shows how the in-coming and out-going waves are excited in form of a signal flow chart as introduced by Hansen. The scalar port wave quantities $v$ and $w$, the transmission vector $\mathbf{T}$, the receiving vector $\mathbf{R}$ and the spherical scattering matrix of the antenna $\mathbf{S}$ are depicted in the figure. The effect of the tissue is modelled by introducing the matrix $\mathbf{U}$
\begin{equation}
    \mathbf{M}_{11} \, \mathbf{b} = \mathbf{a},
\end{equation}
which describes how the out-going SWF are reflected back to the antenna. Note that the out-going coefficients $b_j$ are not only excited by the port ($\mathbf{b}_{\mathrm{port}}=\mathbf{T}v$) but also as back-scattering of the in-coming waves ($\mathbf{b}_{\mathrm{sca}}=\mathbf{S}\,\mathbf{a}$). Therefore the total out-going wave coefficients are defined by 
\begin{equation}
    \mathbf{b} = \mathbf{b}_{\mathrm{port}} + \mathbf{b}_{\mathrm{sca}}.
\end{equation}

The signal flow chart of the original scenario is depicted in the lower part of Fig.~\ref{fig:antenna_refl_screnarios} and shows how the in-coming waves $\mathbf{\hat{a}}$ and the out-going waves $\mathbf{\hat{b}}$ are excited. The coefficients $\mathbf{b}^\prime$ and $\mathbf{a}^\prime$ thereby represent the contributions of the equivalence currents ($\mathbf{J}_{\mathrm{eq}}$, $\mathbf{M}_{\mathrm{eq}}$).

Now where $\mathbf{\hat{E}_0}$ is also decomposed into SWF
\begin{equation}
 \label{eq:e0_recomposed}
    \mathbf{\hat{E}_0} = k \sqrt{\eta} \sum_j \hat{b}_j \mathbf{F}_j^{(4)} + \hat{a}_j \mathbf{F}_j^{(3)},
\end{equation}
the statement \eqref{eq_equivalence} can be formulated in the coefficient space:
\begin{equation}\label{eq_equivalence_coefficients}
    \mathbf{b} = \mathbf{\hat{b}} \hspace{1cm}\mathrm{and}\hspace{1cm} \mathbf{a} = \mathbf{\hat{a}}.
\end{equation}

Using the lower signal flow-graph, the relation between $\mathbf{\hat{b}}$, $\mathbf{\hat{a}}$, $\mathbf{b}^\prime$ and $\mathbf{a}^\prime$ can be derived to:
\begin{equation}
    \mathbf{\hat{b}} = \mathbf{S}^\prime (\mathbf{\hat{a}}-\mathbf{a}^\prime) + \mathbf{b}^\prime.
\end{equation}
This can be further simplified, uing the fact that the free-space spherical scattering matrix $\mathbf{S}^\prime$ is equal to the unity matrix $\mathbf{S}^\prime = \mathbf{I}$, as the origin reflects every in-coming SWF to the corresponding out-going SWF.

As the decomposition can only be approximated by a finite amount of spherical wave functions, the sum in the equations \eqref{eq:e0_original} and \eqref{eq:e0_recomposed} is truncated after an upper bound $N_j$. This is possible, as the series of coefficients of enclosed antennas is converging with respect to the index $j$.

In the following, three approaches to choose the vectors $\mathbf{b}^\prime$ and $\mathbf{a}^\prime$ are discussed, which illustrate the degrees of freedom in the Huygens Equivalence Theorem:
\begin{enumerate}
    \item \label{item_equivalence_love} $\mathbf{b}^\prime = \mathbf{b}, \mathbf{a}^\prime = -\mathbf{a}$: This means that the equivalent currents $\mathbf{J}_{\mathrm{eq}}$ and $\mathbf{M}_{\mathrm{eq}}$ have to be composed by both in-coming and out-going SWF. The recomposed field inside $V$ is $\mathbf{\hat{E}_0}(\mathbf{r}) = \mathbf{0}$. The desired equivalence outside of $V$ is fulfilled. Eqns. \eqref{eq_equivalence} and \eqref{eq_equivalence_coefficients} are fulfilled. This case is equivalent to a special case of the Huygens Equivalence called Love's Equivalence \cite[pp.328]{Balanis2012}. The equivalent currents $\mathbf{J}_{\mathrm{eq}}$ and $\mathbf{M}_{\mathrm{eq}}$ have to be composed by $2N_j$ single mode currents. 
    \item \label{item_equivalence_broken} $\mathbf{b}^\prime = \mathbf{b}, \mathbf{a}^\prime = \mathbf{0}$: In order to reduce the amount of composing currents, a first (naive) approach would be to just leave out the in-coming wave currents in the composition ($\mathbf{a}^\prime = \mathbf{0}$). This means, that the equivalent currents $\mathbf{J}_{\mathrm{eq}}$ and $\mathbf{M}_{\mathrm{eq}}$ only contain out-going SWF, which reduces computational efforts. But unfortunately the eqns. \eqref{eq_equivalence} and  \eqref{eq_equivalence_coefficients} are not fulfilled in this case, which means there is no Huygens Equivalence.
    So this approach would be computationally beneficial, but is actually useless.
    \item \label{item_equivalence_simple} $\mathbf{b}^\prime = \mathbf{b} - \mathbf{a}, \mathbf{a}^\prime = \mathbf{0}$: In order to keep the advantages, of case 2), but to obtain Huygens Equivalence, the $\mathbf{b}^\prime$ are modified. This means that the equivalent currents $\mathbf{J}_{\mathrm{eq}}$ and $\mathbf{M}_{\mathrm{eq}}$ still only contain out-going SWF, which makes it computationally beneficial in contrast to case 1). The recomposed field inside $V$ is $\mathbf{\hat{E}_0}(\mathbf{r}) \neq \mathbf{0}$. However, the desired Huygens Equivalence outside $V$ is fulfilled, meaning eqns. \eqref{eq_equivalence} and \eqref{eq_equivalence_coefficients} are fulfilled. The equivalent currents $\mathbf{J}_{\mathrm{eq}}$ and $\mathbf{M}_{\mathrm{eq}}$ have to be composed by $N_j$ single mode currents.
\end{enumerate}

Cases \ref{item_equivalence_love}) and \ref{item_equivalence_simple}) are suitable for the decomposition in the sense of antenna de-embedding, as both approaches reconstruct the electromagnetic field outside of the equivalent current surface. While case \ref{item_equivalence_love}) is equivalent to the Love's Equivalence Theorem ($\mathbf{J}_{\mathrm{eq}} = \hat{\mathbf{n}} \times \mathbf{H}_0 $ and $\mathbf{M}_{\mathrm{eq}} = -\hat{\mathbf{n}} \times \mathbf{E}_0 $) and might appeal more intuitive in that sense, case \ref{item_equivalence_simple}) has the advantage that only the out-going type of spherical waves are necessary to recompose the electromagnetic field outside.

Even though only the equivalent currents $\mathbf{J}_{\mathrm{eq}}$ and $\mathbf{M}_{\mathrm{eq}}$ currents have to be superimposed by only $N_J$ currents, the composition formula
\begin{equation}
    \label{eq:b_prime}
    \mathbf{b}^\prime = \mathbf{b} - \mathbf{a},
\end{equation}
suggests that still $2N_j = N_j + N_j$ coefficients have to be decomposed in the original domain to calculate $\mathbf{b}^\prime$ from $\mathbf{b}$ and $\mathbf{a}$.

While this would be true, if the formulas \eqref{eq_d4_operator} and \eqref{eq_d3_operator} would be used to obtain the coefficients $a_j$ and $b_j$ separately, there is more efficient way. Whereas \eqref{eq_d4_operator} and \eqref{eq_d3_operator} rely on inner products of the total electric field $\mathbf{E}$ with in- and out-going waves, performing the inner product with a regular wave $\mathbf{F}_j^{(1)}$ (also called standing wave) is also possible. Using \eqref{eq_swf_orthogonality_end} and Table~\ref{tab:a_cc}, the decomposition integral
\begin{equation}
    \label{eq_d1_operator}
    b_j^\prime = b_j - a_j = \frac{2k}{j\sqrt{\eta}} \left<\mathbf{E}, \mathbf{F}_j^{(1)*}\right>,
\end{equation}
can be derived. This surface integral directly calculates the desired coefficients $b_j^\prime$ without first calculating $b_j$ or $a_j$. Using this integral, only $N_j$ coefficients have to be calculated in the original domain therefore as well.

\section{A Note about Numerical Stability}

Even in cases where only out-going waves are expected ($a_j = 0$) it is advisable to use \eqref{eq_d1_operator} in favor of \eqref{eq_d4_operator}. As Santiago et al. \cite{PhysRevB.99.045406} have recently shown in their publication, the decomposition using regular SWF is numerically superior.

\section{Validation of the Approach}

Antenna de-embedding has been prooven to be a useful tool to model complex electromagnetic channels. While Naganawa et al. \cite{Naganawa2017} have shown, that it is suitable to model on-body links, the de-embedding method can also be used also in other approaches, where simulation domains have to be connected. Although the main focus of this paper lies in the theoretical derivation, this section provides illustrates the validity of the method.

\begin{figure}
    \centering
    \includegraphics{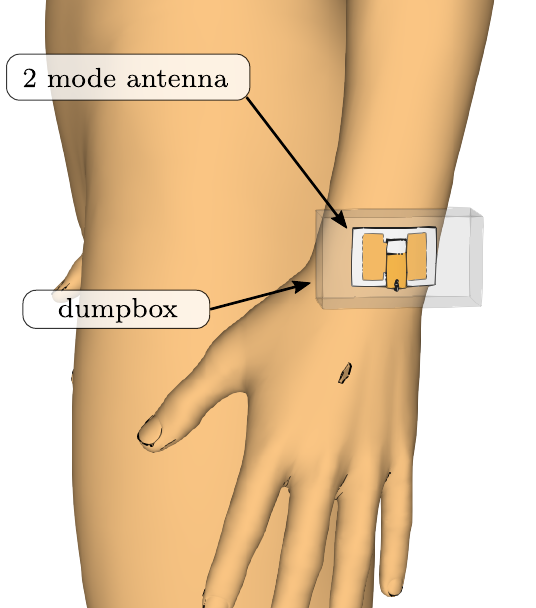}
    \caption{Simulation scenario with dualmode antenna in Empire XPU used as baseline}
    \label{fig:body_antenna}
\end{figure}

\begin{figure}
    \centering
    \includegraphics{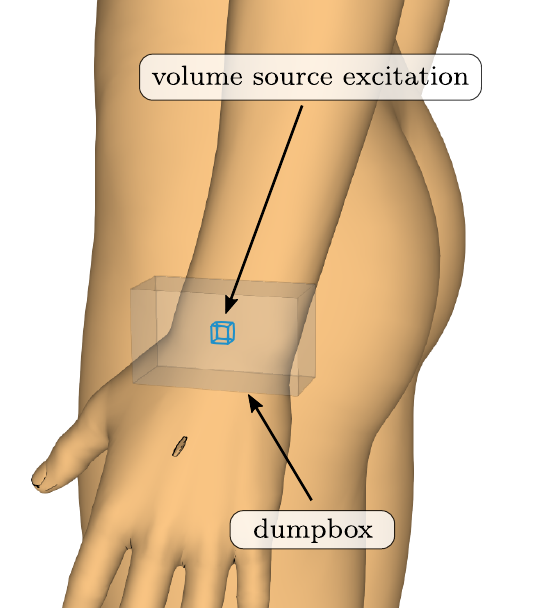}
    \caption{Channel simulation scenario in Empire XPU}
    \label{fig:body_swf}
\end{figure}

\begin{figure}
    \centering
    \includegraphics{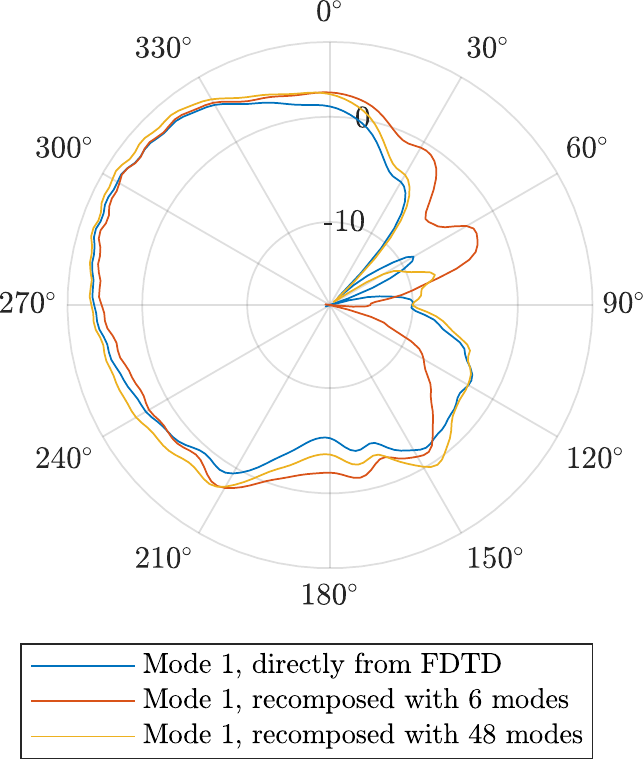}
    \caption{Calculated results for the Off-Body Directivity $D(\theta, \phi = 0^{\circ})$ in dBi of the first mode of the dualmode antenna.}
    \label{fig:mode1}
\end{figure}

\begin{figure}
    \centering
    \includegraphics{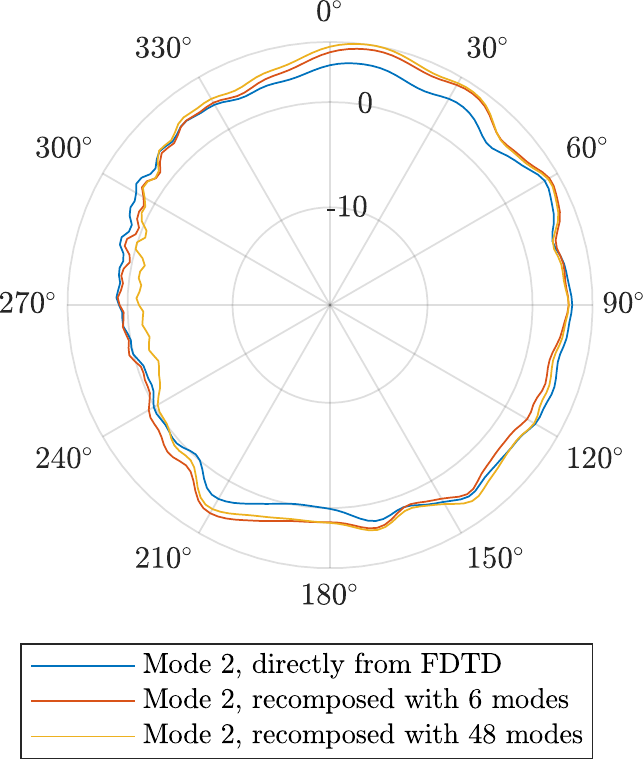}
    \caption{Calculated results for the Off-Body Directivity $D(\theta, \phi = 0^{\circ})$ in dBi of the second mode of the dualmode antenna.}
    \label{fig:mode2}
\end{figure}

The Off-Body-Farfield of a dualmode antenna is calculated by direct simulation and by de-embedding with the first 48 spherical waves. These first 48 modes contain 6 dipole modes, 10 quadrupole modes, 14 hexapole modes and 18 octupole modes:
\begin{enumerate}
    \item \label{step1} A dualmode wristband antenna \cite{DualModeAntenna} is placed at the wrist of a human body model as shown in Fig.~\ref{fig:body_antenna}. A dumpbox enclosing the antenna records and stores the electric and magnetic field values from the FDTD simulations. In the application of the method this simulation would only contain the near environment of the antenna. However, in this example, the entire body is simulated to obtain a baseline and show the validity of the results. Therefore the Off-Body-Farfields of these simulations are also recorded directly.
    \item \label{step2} Fig.~\ref{fig:body_swf} depicts the channel simulation scenario. The antenna is replaced by a volume source excitation box. The simulation is conducted 48 times, while every time another single mode SWF is used as source. The Off-Body-Farfield of every simulation is recorded. Furthermore the dumpbox is used to validate the excited SWF coefficients.
    \item A MATLAB script is used to calculate the SWF coefficients $\mathbf{b}^\prime$ using \eqref{eq_d1_operator} from the stored electromagnetic field values on the dumpbox surfaces from step \ref{step1}). The script then loads the 48 Off-Body-Farfields obtained in step \ref{step2}) and calculates a superposition weighted by the just calculated SWF coefficients. The results are shown in Fig.~\ref{fig:mode1} and \ref{fig:mode2} together with the respective Off-Body-Farfields obtained as baseline directly from \ref{step1}).
\end{enumerate}

The results show, that the de-embedding and superposition leads to the same results as the direct simulation. Already the first six elementary dipole modes are sufficient to approximate the farfield with reasonable accuracy. The results proof and illustrate the validity of the presented de-embedding method.

\section{Summary}

While Naganawa et al. \cite{Naganawa2015} \cite{Naganawa2017} used least mean squares matching to calculate the decomposition coefficients in their publication, we presented a decomposition using surface integrals based on orthogonality in this paper. As this mathematical apparatus is the analytical solution to the least mean squares problem, it provides deeper insights and is more flexible in terms of the decomposition. The decomposition surface integral can be implemented easily and efficiently as post-processing step from FDTD simulation results. Furthermore a way to avoid the need of in-going spherical waves for the correct superposition was derived on a theoretical level. This solidifies the validity of the de-embedding method.


\FloatBarrier



%




\bibliographystyle{IEEEtran}
\bibliography{IEEEabrv,referenzen}

\end{document}